\documentclass[a4paper,11pt]{article}
\usepackage{pos}
\usepackage{lineno}
\newcommand{\pp}{\ensuremath{\mathrm {p\kern-0.05em p}}}
\newcommand{\PbPb}{\ensuremath{\mbox{Pb--Pb}}}
\newcommand{\GeVc}{\ensuremath{\mathrm{GeV}\kern-0.05em/\kern-0.02em c}}
\newcommand{\sqrts}{\ensuremath{\sqrt{s_{\mathrm{NN}}}}}

\newcommand{\pTsub}{\ensuremath{p_{\mathrm{T,subleading}}}}
\newcommand{\pTlead}{\ensuremath{p_{\mathrm{T,leading}}}}
\newcommand{\kT}{\ensuremath{k_{\mathrm{T}}}}

\newcommand{\tg}{\ensuremath{\theta_{\mathrm{g}}}}

\newcommand{\rg}{\ensuremath{R_{\mathrm{g}}}}
\newcommand{\zg}{\ensuremath{z_{\mathrm{g}}}}

\title{Jet substructure measurements in \pp{} and \PbPb{} collisions at $\sqrts=5.02$ TeV with ALICE}
\ShortTitle{Jet substructure measurements in \pp{} and \PbPb{} collisions at $\sqrts=5.02$ TeV with ALICE}

\author[a,b]{James Mulligan for the ALICE Collaboration}
\emailAdd{james.mulligan@berkeley.edu}
\affiliation[a]{Nuclear Science Division, Lawrence Berkeley National Laboratory, Berkeley, California 94720, USA}
\affiliation[b]{Physics Department, University of California, Berkeley, CA 94720, USA}

\abstract{
We report jet substructure measurements in \pp{} and \PbPb{} collisions
at $\sqrts=5.02$ TeV with the ALICE detector.
Charged-particle jets were reconstructed at midrapidity with the 
ALICE tracking detectors using the anti-\kT{} algorithm 
with resolution parameters $R=0.2$ and $R=0.4$.
In \pp{} collisions, the groomed jet momentum fraction, 
$\zg$, and the groomed jet radius, $\tg \equiv \rg/R$, are measured
for the first time using the Dynamical Grooming method. 
Additionally, new systematic measurements of the infrared and collinear (IRC) safe ungroomed jet angularities 
are presented.
In heavy-ion collisions, we measure \zg{} and \tg{} with the Soft Drop grooming algorithm.
The large underlying event in heavy-ion collisions poses
a challenge for the reconstruction of groomed jet observables, 
since fluctuations in the background can cause groomed splittings
to be misidentified.
By using strong grooming conditions to reduce this background,
we report these observables fully corrected for detector effects and background 
fluctuations for the first time, and compare them to several theoretical models.
}

\FullConference{%
  HardProbes2020\\
  1-6 June 2020\\
  Austin, Texas}

\begin{document}
\maketitle

\section{Introduction}

The substructure of jets can be used to study fundamental aspects of QCD
in both \pp{} and \PbPb{} collisions \cite{Larkoski_2020}.
Jet grooming techniques, such as Soft Drop \cite{Larkoski:2014wba} and Dynamical grooming \cite{PhysRevD.101.034004},
reduce non-perturbative effects in \pp{} collisions by selectively removing soft 
large-angle radiation, which allows for well-controlled comparisons of measurements 
to pQCD calculations
\cite{Dasgupta:2013ihk, Larkoski:2015lea, mehtartani2020tagging, Kang:2019prh}.
Grooming techniques have also been applied to heavy-ion collisions,
in order to explore whether jet quenching in the quark-gluon plasma 
modifies the hard substructure of jets
\cite{PhysRevLett.119.112301, Mehtar-Tani2017, Chang:2019nrx, JEWEL2017, Elayavalli2017, Caucal:2019uvr, Ringer_2020, Casalderrey-Solana:2019ubu, Andrews_2020}.
Several measurements of groomed jet observables
have been made in \pp{} and heavy-ion collisions at the LHC and RHIC
\cite{PhysRevD.98.092014, PhysRevD.101.052007, STAR2020, PhysRevLett.120.142302, Acharya:2019djg, Sirunyan2018}.
Ungroomed observables, such as jet angularities \cite{Larkoski_2014}, provide a 
complementary way to study QCD in both \pp{} 
and \PbPb{} \cite{Aad_2012, PhysRevD.98.092014, ang2018} collisions, and offer the ability
to systematically vary the observable definition in a way that is theoretically calculable,
and give sensitivity to the predicted scaling of 
non-perturbative shape functions \cite{Kang_2018, KANG201941}.

In what follows, we reconstruct charged-particle jets at midrapidity with the 
ALICE \cite{aliceDetector} tracking detectors using the anti-\kT{} 
algorithm \cite{antikt, Cacciari:2011ma}
with resolution parameters $R=0.2$ and $R=0.4$.
All presented results are corrected for detector effects (in both \pp{} 
and \PbPb{} collisions) and background fluctuations
(in \PbPb{} collisions) using an iterative unfolding algorithm \cite{DAgostini}.

\section{Jet substructure in proton--proton collisions}

\subsection{Dynamical grooming in proton--proton collisions}

The Dynamical grooming algorithm \cite{PhysRevD.101.034004} identifies a single ``splitting'' by 
re-clustering the constituents of a jet with the Cambridge-Aachen algorithm \cite{Dokshitzer_1997}, 
and traversing the primary Lund plane \cite{Dreyer_2018} to identify the splitting that maximizes:
$
z_i (1-z_i)p_{\rm{T},i} \left( \frac{\Delta R_i}{R} \right)^a,
$
where $z_i$ is the longitudinal momentum fraction of the $i^{\rm{th}}$ splitting,
$\Delta R_i$ is the rapidity-azimuth ($y,\varphi$) separation of the daughters,
and $a$ is a continuous free parameter. 
Since the grooming condition defines a maximum rather than an explicit cut 
(as in the case of Soft Drop), every jet will always return a tagged splitting.
We focus on the two kinematic observables that characterize the splitting: 
the groomed jet radius, $\tg \equiv \rg/R \equiv \sqrt{\Delta y ^2 + \Delta \varphi ^2}/R$,
and the groomed momentum fraction, $\zg \equiv \pTsub / (\pTlead + \pTsub)$.

\begin{figure}[!ht]
\centering{}
\includegraphics[scale=0.34]{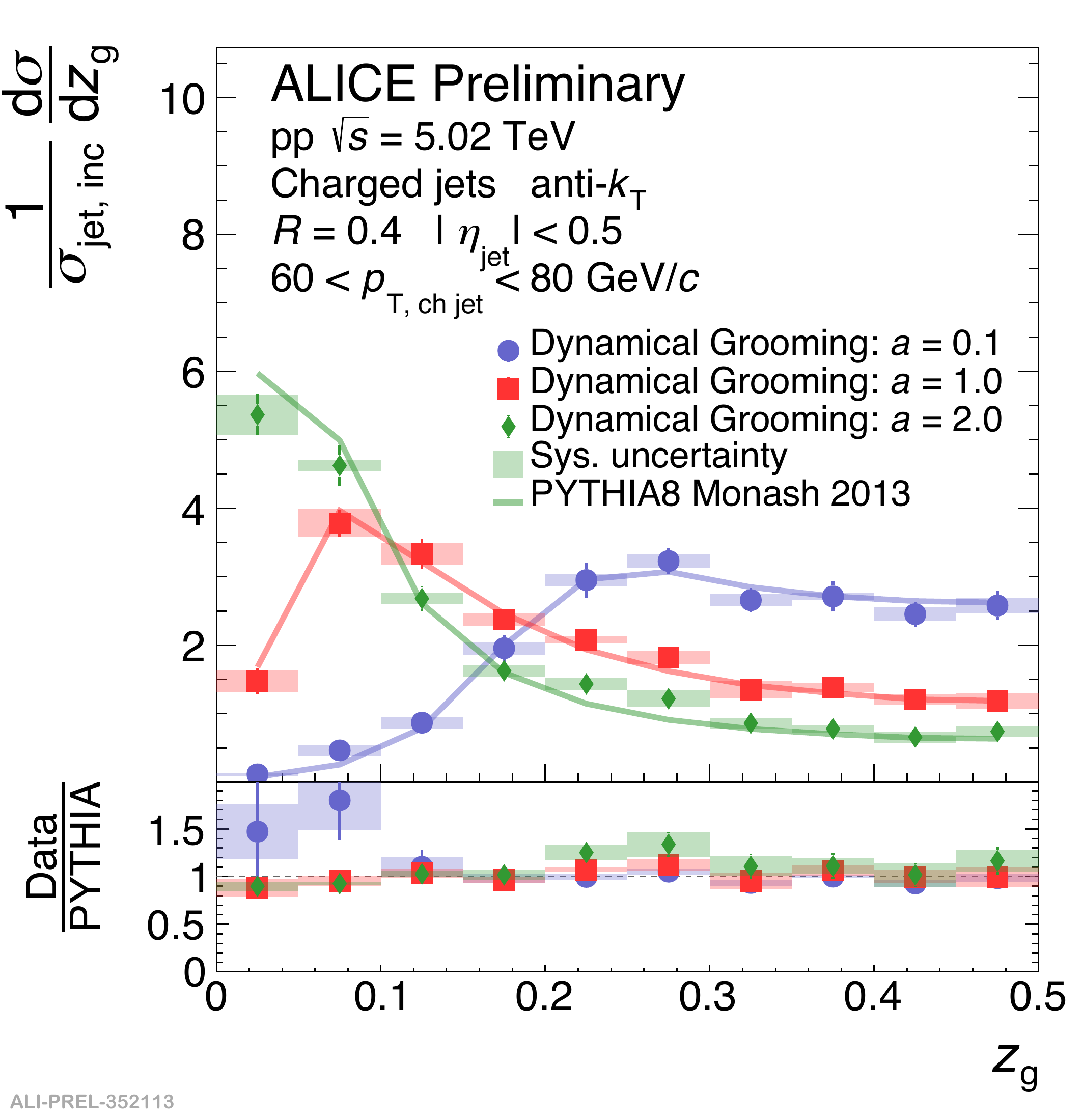}
\includegraphics[scale=0.34]{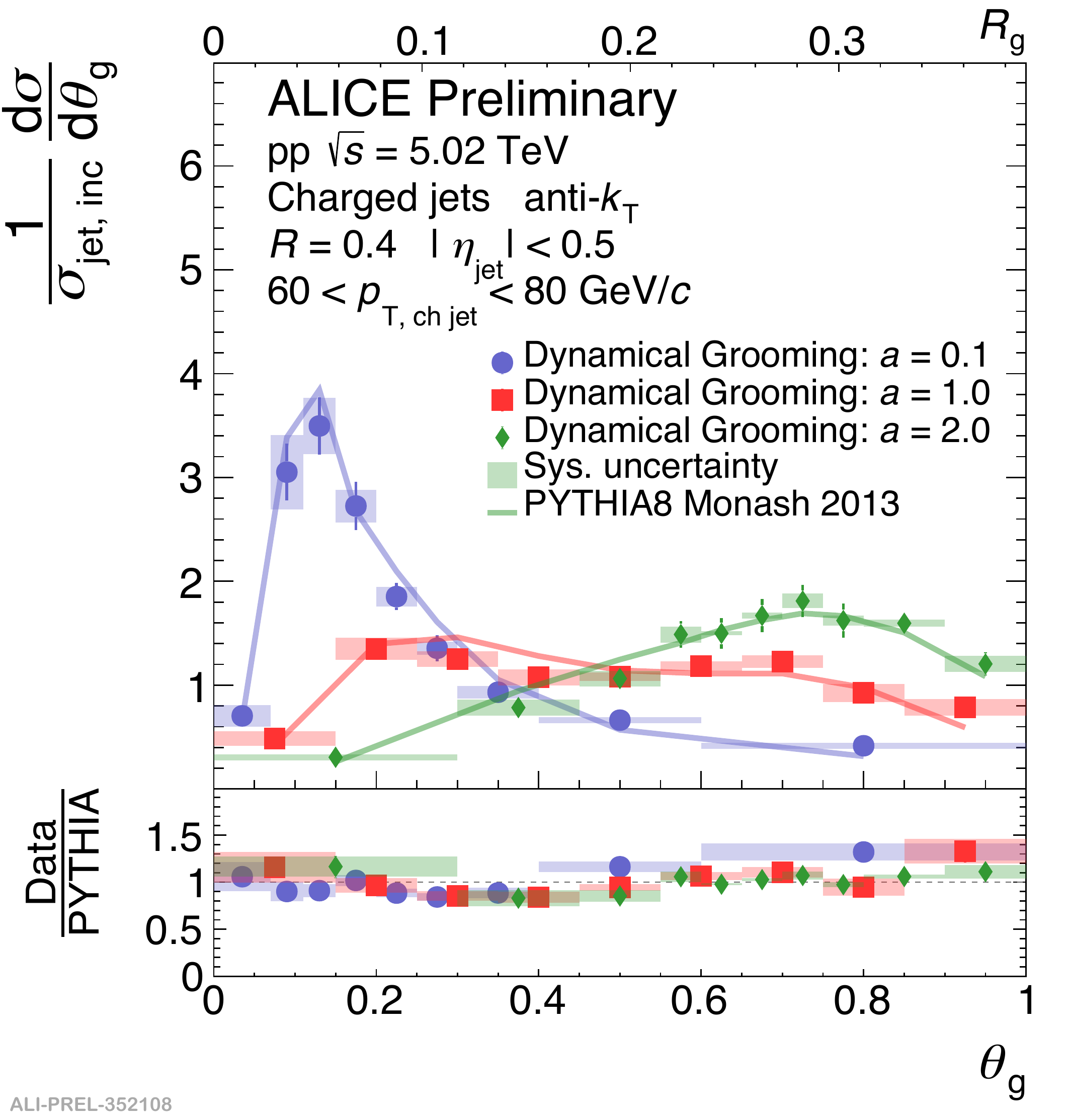}
\caption{Measurements of \zg{} (left) and \tg{} (right) in \pp{} collisions with
Dynamical grooming \cite{PhysRevD.101.034004} for three values of the
grooming parameter $a$, along with comparison to PYTHIA Monash 2013 \cite{pythia}.}
\label{fig:dg}
\end{figure}

Figure \ref{fig:dg} shows the \zg{} and \tg{} distributions in \pp{} collisions
for several values of the grooming parameter $a$. 
For small values of $a$, the grooming condition favors splittings with symmetric
longitudinal momentum, which is reflected in the distributions skewing towards
large-\zg{} and small-\tg. As $a$ increases, the grooming condition favors 
splittings with large angular separation, which is reflected in the distributions 
skewing towards small-\zg{} and large-\tg.
The results are compared to PYTHIA \cite{pythia}, 
which describes the data well.

\subsection{Ungroomed jet angularities in proton-proton collisions}

\begin{figure}[!b]
\centering{}
\includegraphics[scale=0.34]{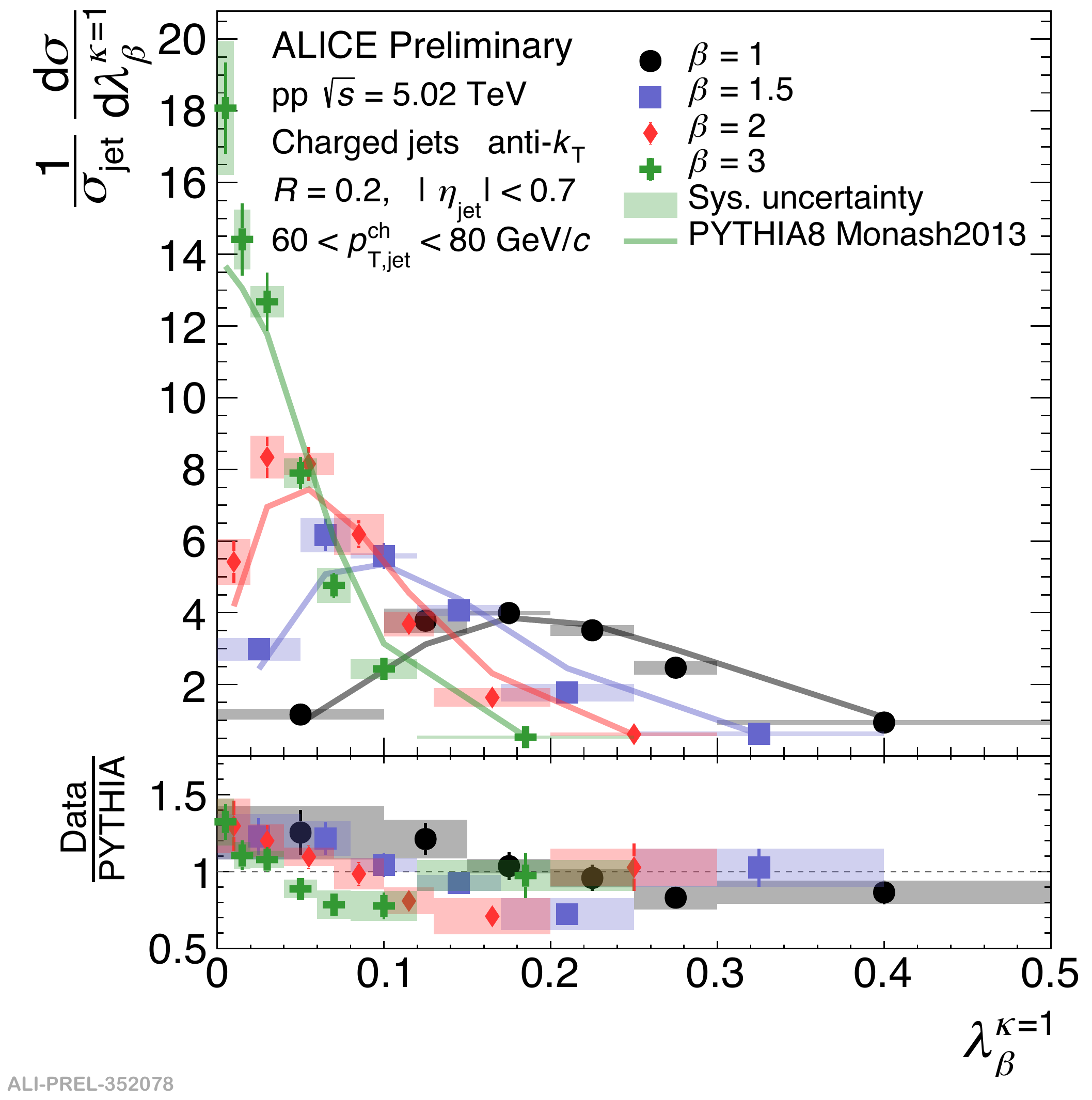}
\includegraphics[scale=0.34]{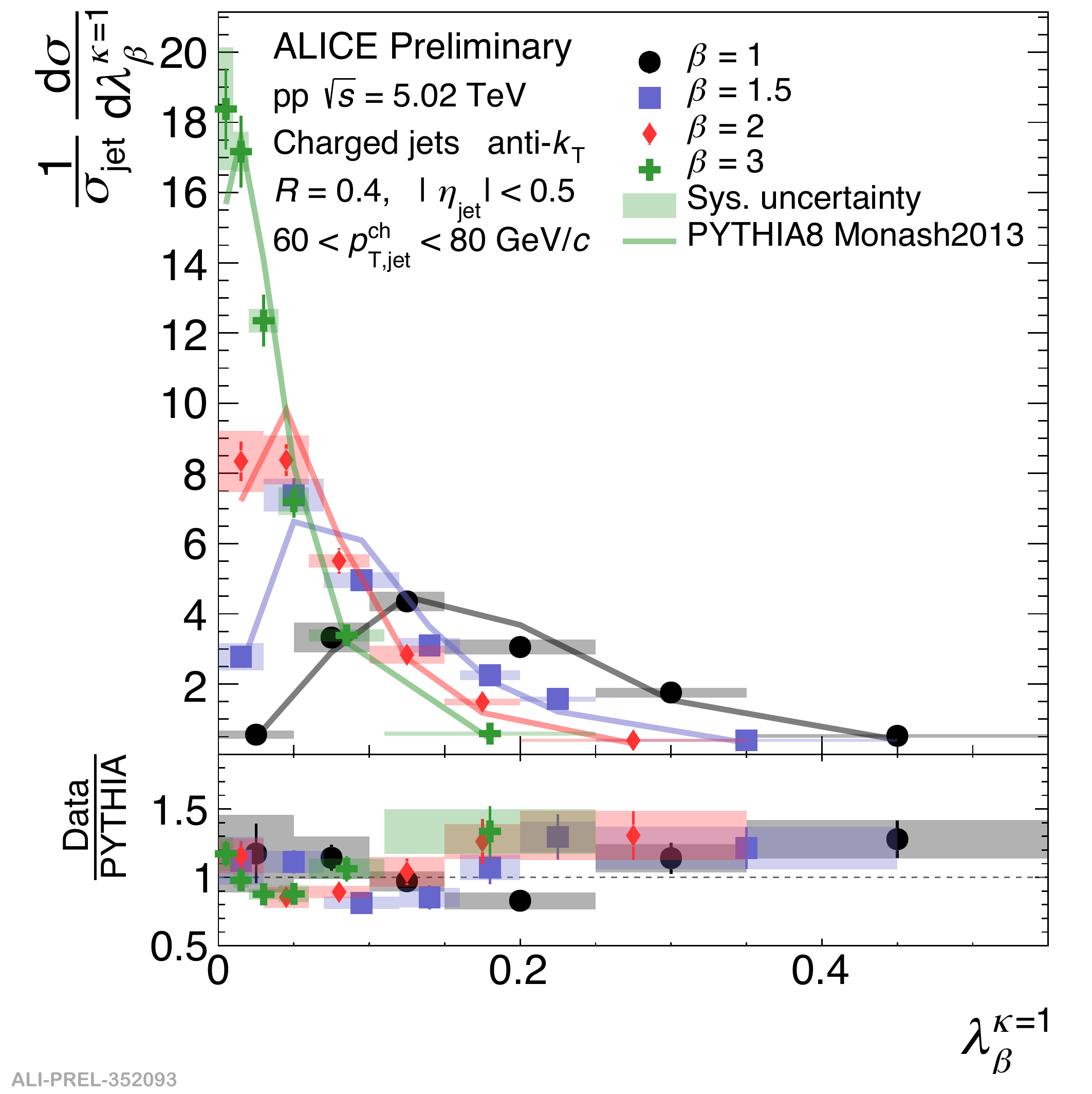}
\caption{Measurements of jet angularities $\lambda_\beta^{\kappa=1}$ in \pp{} collisions with
for $R=0.2$ (left) and $R=0.4$ (right) for four values of the
continuous parameter $\beta$, along with comparison to PYTHIA Monash 2013 \cite{pythia}.}
\label{fig:ang}
\end{figure}

The class of IRC-safe jet angularities \cite{Larkoski_2014} are defined
as 
\begin{equation} \label{ang_eqn}
\lambda_\beta^{\kappa} =\sum\limits_{i \in \text{jet}}
\bigg( \frac{p_{\text{T},i}}{p_{\text{T, jet}}} \bigg)^{\kappa}
\bigg( \frac{\Delta R_i}{R} \bigg)^\beta 
\end{equation}
for $\kappa=1$ and $\beta>0$. 

Figure \ref{fig:ang} shows the $\lambda_\beta^{\kappa=1}$ distributions in \pp{} collisions
for $R=0.2$ (left) and $R=0.4$ (right) for several values of $\beta$. 
As $\beta$ increases, the distributions skew towards small $\lambda_\beta^{\kappa=1}$, 
since $\Delta R_i/R$ is smaller than unity. 
For larger $R$, the distributions are narrower
than for smaller $R$, 
as expected due to the collinear nature of jet fragmentation.
The results are compared to PYTHIA \cite{pythia}, 
which describes the data reasonably well but with some deviations to be further explored.

\section{Jet substructure in heavy-ion collisions}

\begin{figure}[!ht]
\centering{}
\includegraphics[scale=0.3]{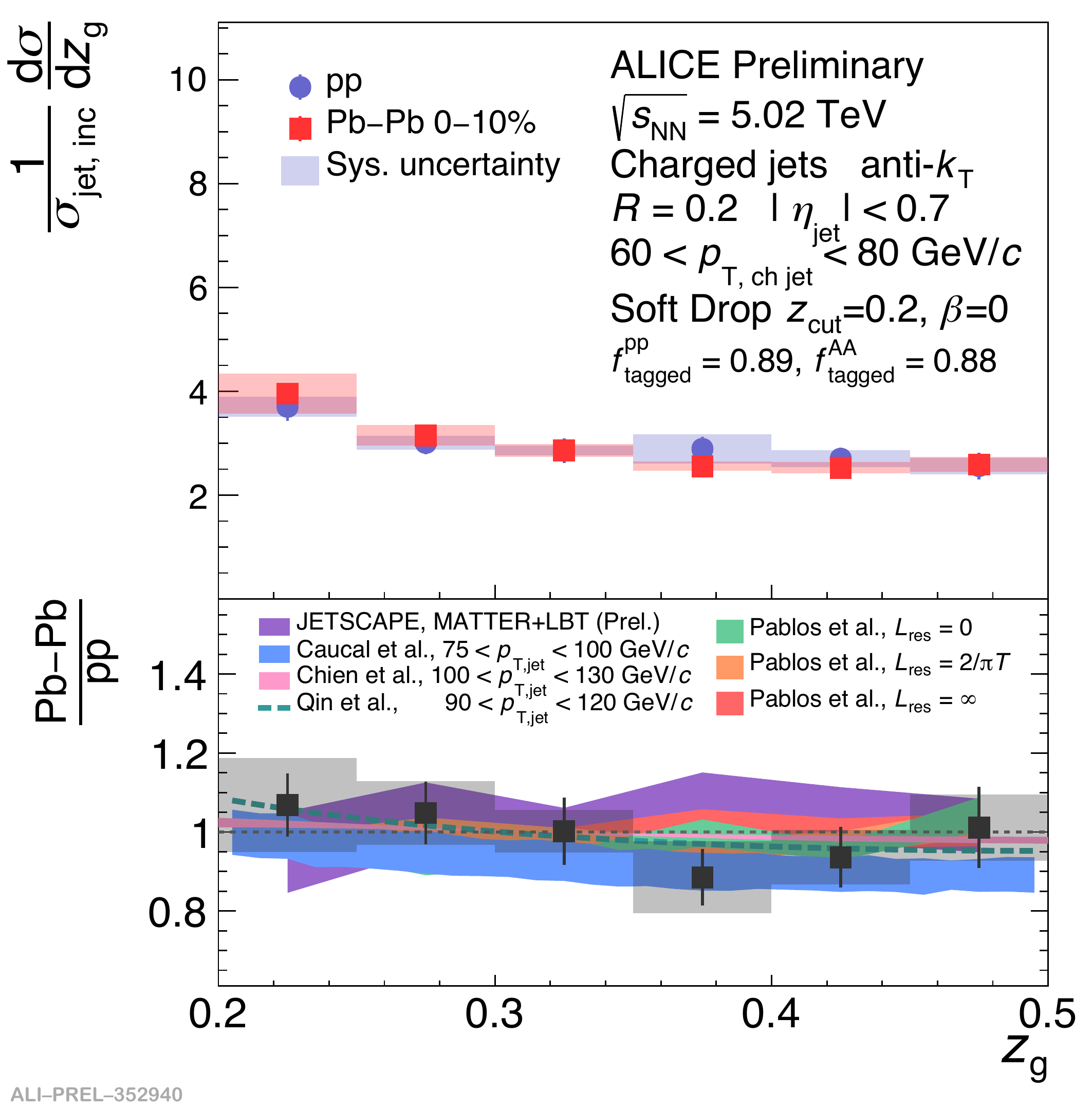}
\includegraphics[scale=0.3]{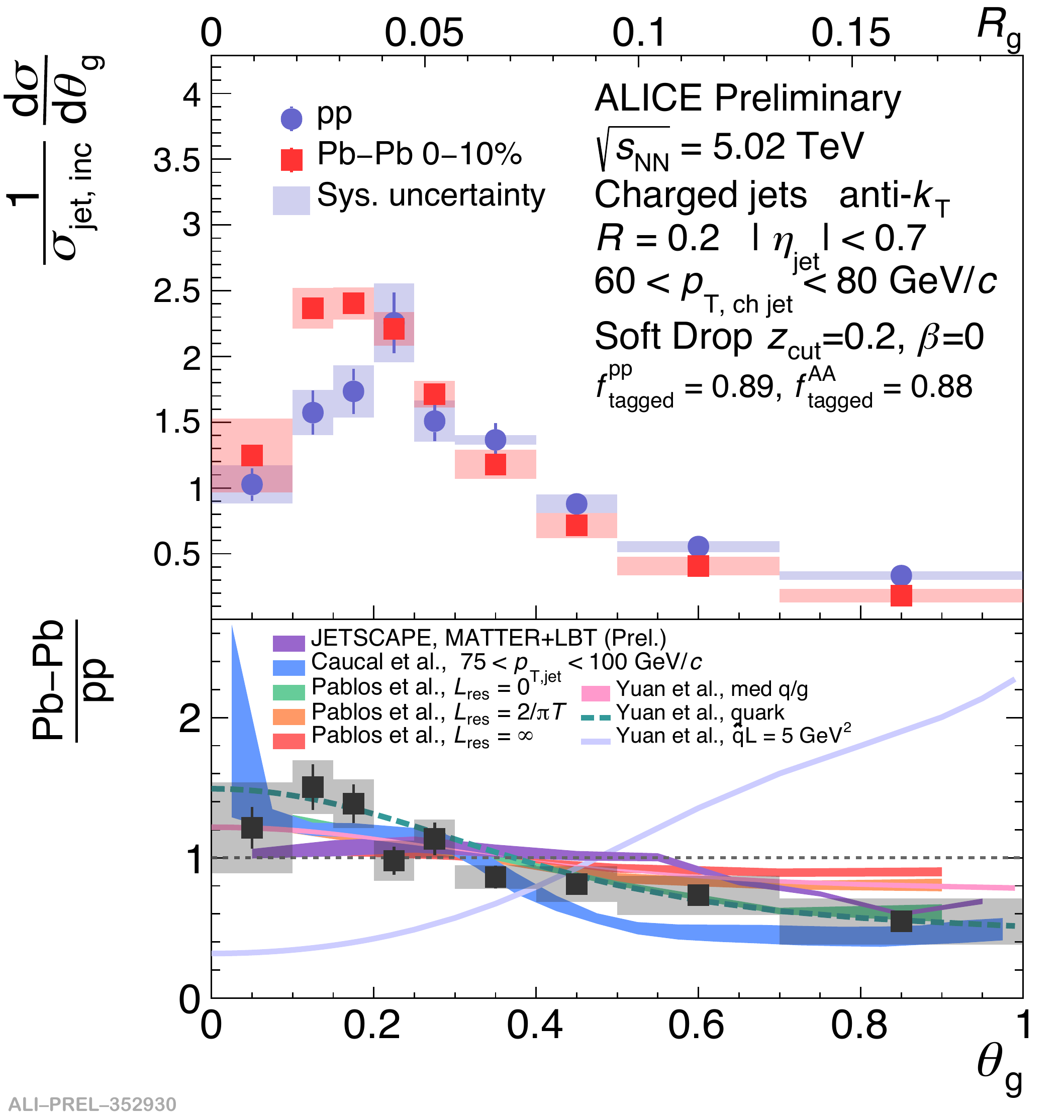}
\caption{Measurements of \zg{} (left) and \tg{} (right) in 0--10\% central \PbPb{} collisions compared to \pp{} collisions for $R=0.2$, along with comparison to several theoretical models \cite{ALICE-PUBLIC-2020-006}.}
\label{fig:sd-central}
\end{figure}

\begin{figure}[!ht]
\begin{center}
\includegraphics[scale=0.09]{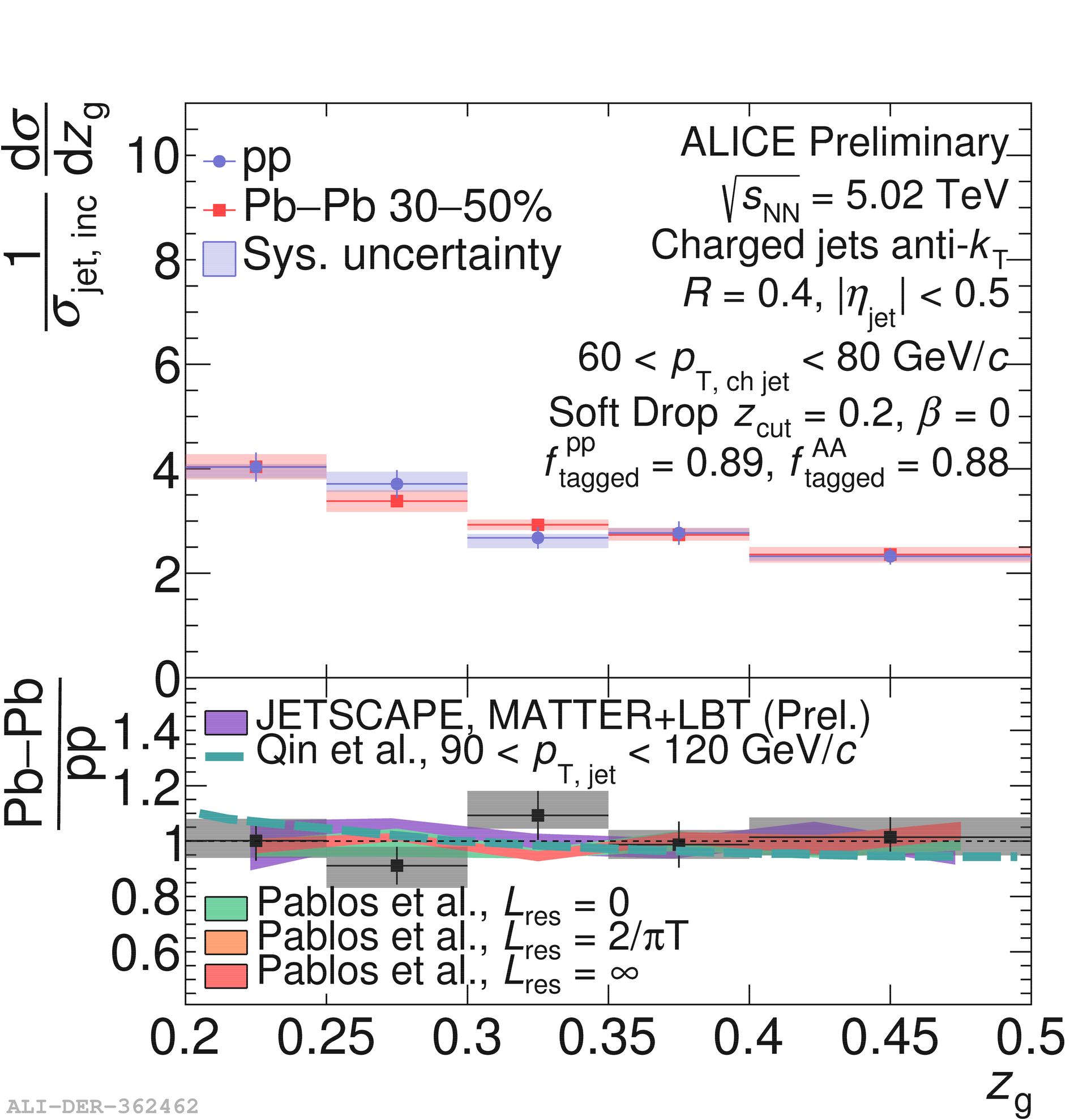}
\includegraphics[scale=0.09]{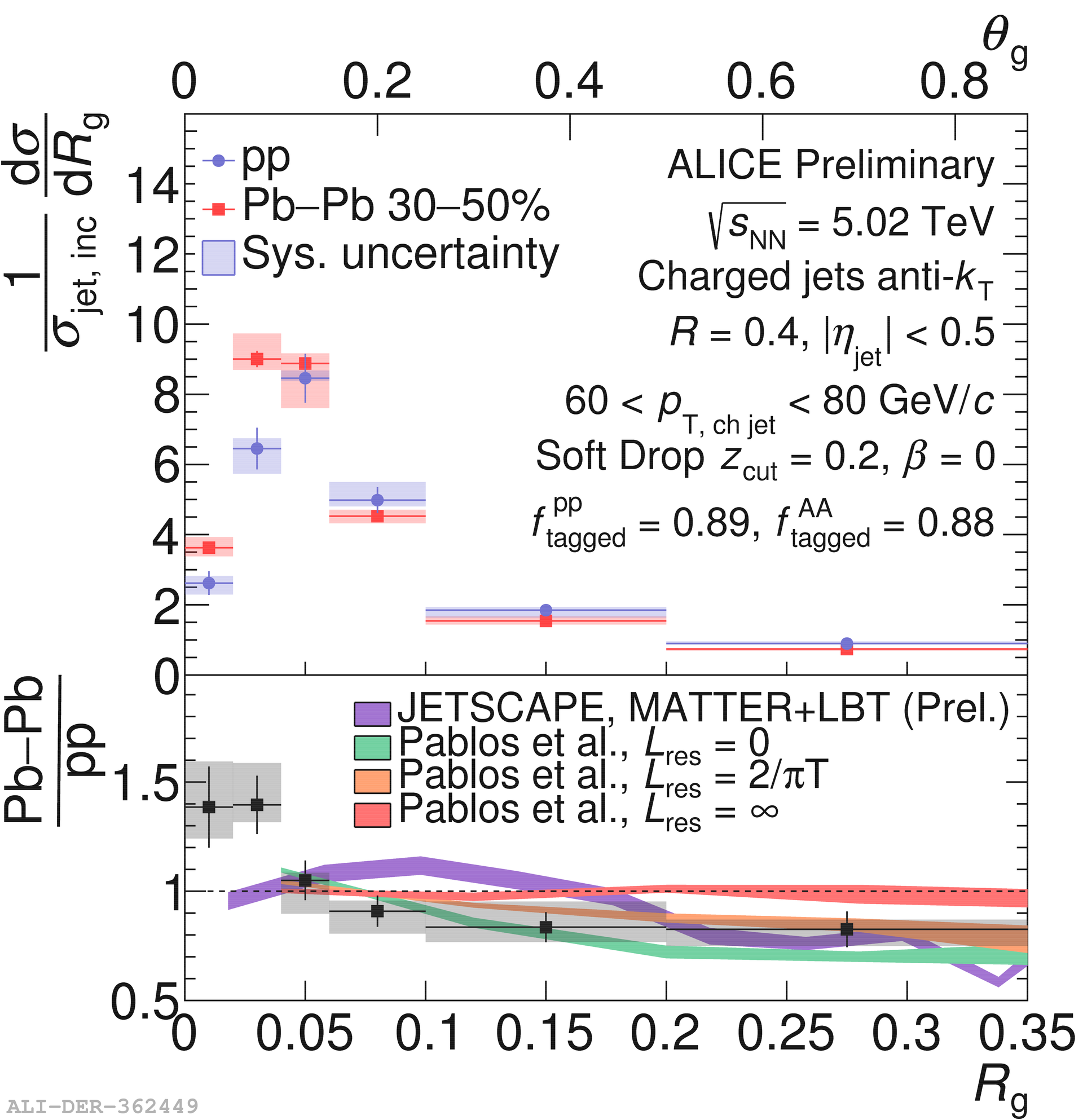}
\caption{Measurements of \zg{} (left) and \rg{} (right) in 30--50\% central \PbPb{} collisions compared to \pp{} collisions for $R=0.4$, along with comparison to several theoretical models \cite{ALICE-PUBLIC-2020-006}.}
\label{fig:sd-semicent}
\end{center}
\end{figure}

In heavy-ion collisions, the large underlying event poses
a challenge for the reconstruction of groomed jet observables, 
since fluctuations in the background can cause groomed splittings
to be misidentified \cite{mulligan2020identifying}.
We present measurements \cite{ALICE-PUBLIC-2020-006} of \zg{} and \tg{} with the Soft Drop
grooming algorithm that are fully corrected for detector effects and background 
fluctuations, leveraging stronger grooming conditions 
than in previous measurements.
Figures \ref{fig:sd-central} and \ref{fig:sd-semicent} show these measurements
in \PbPb{} collisions together with their comparison to those
from \pp{} collisions, for
central (0--10\%) and  semi-central (30--50\%) \PbPb{} collisions, respectively.

We find that the \zg{} distributions in \PbPb{} collisions are consistent with those 
in \pp{} collisions, whereas a significant narrowing of the \tg{} distributions 
in \PbPb{} collisions relative to  \pp{} collisions is observed.
These measurements are compared to a variety of jet quenching models:
JETSCAPE \cite{Putschke:2019yrg, LBT, Majumder_2013},
Caucal et al. \cite{Caucal:2019uvr, Caucal_2018}, 
Chien et al. \cite{PhysRevLett.119.112301},
Qin et al. \cite{Chang:2019nrx},
Pablos et al. \cite{HybridModel, HybridModelResolution, Casalderrey-Solana:2019ubu}, and
Yuan et al. \cite{Ringer_2020, Qiu:2019sfj}.
All models considered are consistent with the \zg{} measurements.
Many of the models capture the narrowing effect observed in the \tg{} distributions,  
although with quantitative differences. 
This behavior is consistent with models implementing an incoherent interaction of the 
jet shower constituents with the medium, but also consistent with
medium-modified ``quark/gluon'' fractions  with fully coherent energy loss.
By isolating the theoretically well-controlled hard substructure of jets, these measurements
provide direct connection to specific jet quenching physics mechanisms, 
and offer the opportunity for future measurements to definitively disentangle them.

{
\scriptsize
\bibliographystyle{JHEP}
\bibliography{main.bib}
}

\end{document}